\documentclass{elsart1p}
\usepackage{graphicx}
\usepackage{amssymb}
\newcommand{\bra}{\langle}
\newcommand{\ket}{\rangle}
\newcommand{\bs}[1]{\mbox {\boldmath $#1$}}
\begin{document}
\begin{frontmatter}
%
%
%
\title{Weak reactions with light nuclei - $^6$He $\beta$-decay as a test case
  for the nuclear weak current}
\author{Doron Gazit\thanksref{doe}}
\ead{doron.gazit@mail.huji.ac.il}
\address{Institute for Nuclear Theory, University of Washington, 
Box 351550, 98195 Seattle, WA, USA}
\author{Sergey Vaintraub, Nir Barnea\thanksref{isf}}
\ead{sergey.vaintraub@mail.huji.ac.il}
\thanks[doe]{work supported by DOE grant number DE- 
FG02-00ER41132}
\thanks[isf]{work supported by the ISRAEL SCIENCE FOUNDATION (grant no. 361/05)}
\address{Racah Institute of Physics, The Hebrew University, Jerusalem, 91904, Israel}
\ead{nir@phys.huji.ac.il}
%
%
%
\begin{abstract}
We present a microscopic calculation of the $^6$He $\beta$-decay into the ground
state of $^6$Li. To this end we use the impulse approximation to describe the
nuclear weak current. The ground state wave
functions are obtained from the solution of the nuclear $6$-body problem.
The nucleon-nucleon interaction is described via the J-matrix inverse scattering 
potential (JISP), and the nuclear problem is solved using the
hyperspherical-harmonics approach. This approach results in numerical accuracy
of about 2 per mil in the transition matrix element. Bearing in mind that the
contribution of meson-exchange currents to the transition matrix element is
about $5\%$, these results pave the way 
for  
accurate estimation of their effect.
\end{abstract}
\end{frontmatter}
%
\section{Introduction}
\label{Int}
The nuclear weak current contains a well established leading 1-body terms,
which at low energy 
are known as the Fermi (F) and Gamow-Teller (GT) operators, and subleading  
two- and many-body currents, generally known as meson-exchange currents (MEC),
which are a subject of vast contemporary research.
The lightest nucleus that undergoes a $\beta$-decay is the
triton. However, the theory of nuclear weak interaction cannot be checked in
the triton since its 
half-life is used to adjust a free parameter in the axial MEC. Therefore, the
lightest nucleus that can provide a test to the theory is 
$^6$He. $^6$He ($\rm{J}^\pi=0^{+}$) is an unstable nucleus, which undergoes
a $\beta$ decay with a half-life $\tau_{1/2}=806.7 \pm
1.5\,\rm{msec}$ to the ground state of $^6$Li ($\rm{J}^\pi=1^{+})$
\cite{AjzenbergSelove:1988ec}.

So far, a microscopic calculation of $^6$He from its
nucleonic degrees of freedom, failed to reproduce the $\beta$-decay
rate. A comprehensive study, accomplished by Schiavilla and Wiringa
\cite{SchiavillaPhysRevC2002}, has used the realistic Argonne $v18$
(AV18) nucleon-nucleon (NN) potential, combined with the Urbana-IX (UIX)
three-nucleon-force (3NF), to derive the nuclear wave functions, through the
variational Monte-Carlo approach (VMC). The model used for the nuclear weak
axial 
current includes one- and two-body operators. The two-body currents are
phenomenological, with the strength of the leading two-body term -- associated
with $\Delta$-isobar excitation of the nucleon -- adjusted to reproduce the
GT matrix element in triton $\beta$-decay. The calculated
half-life of $^6$He is over-predicts the measured one by about 8\%. An
unexpected result of the calculation, was that two-body currents lead to a
$1.7\%$ increase in the value of the Gamow-Teller matrix element of $^6$He,
thus worsening the comparison with experiment. The authors of this paper have
presumed that the origin of this discrepancy is either the in approximate
character of the VMC wave functions, or in the model
of the weak nuclear current.  

The $\beta$ decay rate of $^6$He is proportional to the square of the
GT matrix element. The difference between
the one-body contribution to the $^6$He-$^6$Li GT matrix element and the
experimental value, $2.173\pm 0.002$, is of the order of few percent. A
result which on the one hand is very satisfying, but on the other hand implies
that numerical accuracy at a per mil level is required to assign the
$^6$He $\beta$-decay as a test case for validating the MEC model.   
 
The current contribution is dedicated to explore exactly that point. Having in
mind the required level of convergence we use the JISP16 potential
\cite{JISP16}. The JISP16 NN potential
utilizes the J-matrix inverse scattering technique to construct a soft nuclear
potential, formulated in the harmonic oscillator basis, that by construction
reproduces the NN phase shifts up to pion threshold and the binding energies of
the light nuclei $A\leq 4$.    

We use the Hyperspherical-Harmonics (HH) expansion to solve the Schr\"{o}dinger
equation. 
The HH functions constitute a general basis
for expanding the wave functions of an $A$-body
system~\cite{A89,F83}. 
In the HH method, the translational invariant wave-function is written as
\begin{equation}  
 \Psi = \sum_{[K]n} C_{[K]n}R_{n}(\rho){\cal Y}_{[K]}(\Omega,s_i,t_i)
\end{equation}
where $\rho$ is the hyperradius, and  $R_{n}(\rho)$ are a complete set of
basis functions symmetric under particle permutations as
$\rho^2=\frac{1}{2A}\sum_{i,j}(\bs{r}_i-\bs{r}_j)^2$. The hyperangle,
$\Omega$, is a set of $3A-4$ angles, and ${\cal Y}_{[K]}(\Omega,s_i,t_i)$ are a
complete set of antisymmetric basis functions in the Hilbert space of
spin, isospin and hyperangles. The functions ${\cal Y}_{[K]}(\Omega,s_i,t_i)$
are characterized by a set 
of quantum numbers $[K]$ \cite{Barnea97,Barnea99}, and
possess definite angular momentum, isospin, and 
parity quantum numbers.
They are
eigenfunctions of the hyperspherical, or generalized, angular momentum
operator $\hat{K}^2$, $\hat{K}^2{\cal Y}_{[K]}(\Omega,s_i,t_i)=K(K+3A-5){\cal
  Y}_{[K]}(\Omega,s_i,t_i)$.

Our results for the ground state properties of $^{6}$He and $^{6}$Li, are
presented in table \ref{tab6}.  
The energies and rms
matter radii are given as a function of $K_{max}$ -- the limiting value of $K$ in
the HH expansion. As we have used the bare interaction in our calculations the
results are variational.
Using the formula $E(K_{max})=E_{\infty}+A/K_{max}^{\alpha}$
for $K_{max}\geq 8$
we have extrapolated the binding energies to the limit
$K_{max}\longrightarrow \infty$. It can be seen that the extrapolated binding energies are 
rather close to the experimental values. In the last column of the 
table we present our $^6$He-$^6$Li GT transition matrix element in the impulse
approximation, i.e. at the 1-body level. It can be seen
that
the convergence pattern of the matrix element is not regular. Extrapolating
its value using the expression
$\rm{GT}(K_{max})=\rm{GT}_{\infty}+B/K_{max}^{\beta}$
for $K_{max}\geq 6$, we get $\rm{GT}_{\infty}=2.228(3)$.
This result is in accordance with the values $\rm{GT}=2.28$ for AV8'/TM'(99) and
$\rm{GT}=2.30$ for AV8' obtained by Navratil and Ormand \cite{Navratil03},
$\rm{GT}=2.28$ for the N$^3$LO NN-force of Navratil and Caurier
\cite{Navratil04}, $\rm{GT}=2.25$ for AV18/UIX of Schiavilla and Wiringa
\cite{SchiavillaPhysRevC2002}, and $\rm{GT}=2.16-2.21$ for AV18/IL2 by
Pervin {\it et} al. \cite{Pervin2007}. Moreover, it can be seen that our
accuracy estimating the GT matrix element is in the level of few per mil.
Such an accuracy enables us to disentangle numerics from physics and utilizes
the $^{6}$He $\beta$-decay as a test ground for axial MEC models. 

\begin{table}[h]
 \centering
 \label{tab6}
 \caption{The $^{6}$He, $^{6}$Li binding energies, rms matter radius, and GT
   matrix element in the impulse approximation. Results for the JISP16 NN
   interaction.} 
   \begin{tabular}[h]{rccccc}
   \\   \hline  \hline 
\hspace{3mm}  $K_{max}$       \hspace{1mm} &  
\hspace{3mm}  B.E.($^{6}$He)  \hspace{3mm} & 
\hspace{3mm}  B.E.($^{6}$Li)  \hspace{3mm} & 
\hspace{3mm}  $r_{rms}$($^{6}$He) \hspace{3mm}  & 
\hspace{3mm}  $r_{rms}$($^{6}$Li) \hspace{3mm}  &  
\hspace{3mm} $\bra \rm{GT} \ket$  \hspace{3mm}\\
                  \hline 
    4   \hspace{2mm} & 18.367 & 19.392 & 1.840 & 1.859 & 2.263 \\
    6   \hspace{2mm} & 24.103 & 26.124 & 1.902 & 1.909 & 2.247 \\
    8   \hspace{2mm} & 26.392 & 28.854 & 1.979 & 1.984 & 2.234 \\
   10   \hspace{2mm} & 27.560 & 30.156 & 2.051 & 2.051 & 2.232 \\
   12   \hspace{2mm} & 28.112 & 30.797 & 2.112 & 2.110 & 2.229 \\
   14   \hspace{2mm} & 28.424 &        &       &       &       \\
$\infty$\hspace{3mm} & 29.041 & 31.923 & 2.165 &       &       \\
   \hline 
Shirokov {\it et} al. \cite{JISP16}  
                     & 28.32(28) &  31.00(31) &       &       \\ 
   \hline 
  Exp. & 29.269 & 31.995 & 2.18  & 2.09  & 2.170 \\
                \hline 
                \hline\\
                \end{tabular}
\end{table}

%
%
%

%
\end{document}